\documentclass[pdftex,pagesize=auto,paper=letter,9pt,abstract=on,twocolumn,DIV=calc]{scrartcl} \usepackage{amsmath,mathtools,amsthm} \usepackage{amsfonts} \usepackage{graphicx} \usepackage{algorithm} \usepackage{algorithmicx} \usepackage[noend]{algpseudocode} \usepackage[pdftex]{xcolor} \usepackage{hyperref} \usepackage{booktabs} \usepackage{pgffor} \usepackage{cite} \usepackage{flushend} \usepackage[format=plain,labelformat={simple},labelsep={period},labelfont={bf}]{caption} \usepackage[captionskip=1em,font={small}]{subfig} \usepackage{microtype} \usepackage{lmodern} \usepackage[T1]{fontenc}

\newtheorem{theorem}{Theorem}[section]

\newcommand{\ignore}[1]{} \newcommand{\notinproc}[1]{}

\newcommand{\HC}{\mathit{HC}} \newcommand{\comment}[1]{}

\definecolor{darkgreen}{RGB}{0,100,0} \definecolor{orange}{RGB}{255,80,0}     \newcommand{\Xcomment}[1]{} \newcommand{\instance}[1]{#1}

\newcommand{\increase}[2]{\ensuremath{#1 \, {\overset{+}{\gets}}\, #2}} 

\title{Computing Classic Closeness Centrality, at Scale} \date{August 2014} \author{ EDITH COHEN\\Microsoft Research\\editco@microsoft.com \and DANIEL DELLING\\Microsoft Research\\dadellin@microsoft.com \and THOMAS PAJOR\\Microsoft Research\\tpajor@microsoft.com \and RENATO F.~WERNECK\\Microsoft Research\\renatow@microsoft.com }

\begin{document}

\maketitle

\begin{abstract} Closeness centrality, first considered by Bavelas~(1948), is an importance measure of a node in a network which is based on the distances from the node to all other nodes. The classic definition, proposed by Bavelas~(1950), Beauchamp~(1965), and Sabidussi~(1966), is~(the inverse of) the average distance to all other nodes.

We propose the first highly scalable (near linear-time processing and linear space overhead) algorithm for estimating, within a small relative error, the classic closeness centralities of all nodes in the graph. Our algorithm applies to undirected graphs, as well as for centrality computed with respect to round-trip distances in directed graphs.

For directed graphs, we also propose an efficient algorithm that approximates generalizations of classic closeness centrality to outbound and inbound centralities. Although it does not provide worst-case theoretical approximation guarantees, it is designed to perform well on real networks.

We perform extensive experiments on large networks, demonstrating high scalability and accuracy. \end{abstract}

\section{Introduction}

Closeness centrality is a structural measure of the importance of a node in a network, which is based on the ensemble of its distances to all other nodes. It captures the basic intuition that the closer a node is to all other nodes, the more important it is. Structural centrality in the context of social graphs was first considered in 1948 by Bavelas~\cite{Bavelas:HumanOrg1948}.

The classic definition measures the closeness centrality of a node as the inverse of the average distance from it and was proposed by Bavelas~\cite{Bavelas:Acous1950}, Beauchamp~\cite{Beauchamp:BS1965}, and Sabidussi~\cite{Sabidussi:psychometrika1966}. On a graph $G = (V,E)$ with $|V|=n$ nodes, the centrality of $v$ is formally defined by \begin{equation} \label{Bclosenesscdef} B^{-1}(v) = (n-1)/\sum_{u \in V} d_{vu}, \end{equation} where $d_{vu}$ is the shortest-path distance between $v$ and $u$ in $G$. This textbook definition is also referred to as {\em Bavelas closeness centrality} or as the {\em Sabidussi Index}~\cite{Freeman:sociometry1977,Freeman:sn1979,wassermansocialnetworksbook}.

The classic closeness centrality of a node $v$ can be computed exactly using a single-source shortest paths computation (such as Dijkstra's algorithm). In general, however, we are interested not only in the centrality of a particular node, but rather in the set of all centrality values. This is the case when centrality values are used to obtain a relative ranking of the nodes. Beyond that, the distribution of centralities captures important characteristics of a social network, such as its {\em centralization}~\cite{Freeman:sn1979,wassermansocialnetworksbook}.

When we would like to perform many centrality queries (in particular when we are interested in centrality values for all nodes) on graphs with billions of edges, such as large social networks and Web crawl graphs, the exact algorithms do not scale. Instead, we are looking for scalable computation of approximate values, with small relative error.

The node with maximum classic closeness centrality is known as the 1-median of the network. A near-linear-time algorithm for finding an approximate 1-median was proposed by Indyk and Thorup \cite{Indyk:stoc1999,Thorup:icalp2001}. Their algorithm samples $k$ nodes at random and performs Dijkstra's algorithm from each sampled node. They show that the node with minimum sum of distances to sampled nodes is with high probability an approximate 1-median of the network. The same sampling approach was also used to estimate the centrality values of all nodes~\cite{EW_centrality:SODA2001} and to identify the top $k$ centralities~\cite{OkamotoCL:FAW2008}. When the distance distribution is heavy-tailed, however, the sample average is a very poor estimator of the average distance: The few very distant nodes that dominate the average distance are likely to be all excluded from the sample $C$, resulting in a large expected error for almost all nodes.

\subsubsection*{Contributions} We present the first near-linear-time algorithm for estimating, with a small relative error, the classic closeness centralities of all nodes. Our algorithm provides probabilistic guarantees that hold for all instances and for all nodes.

\begin{figure*}[htbp] \centering\subfloat[Exact]{\includegraphics[width=0.22\textwidth]{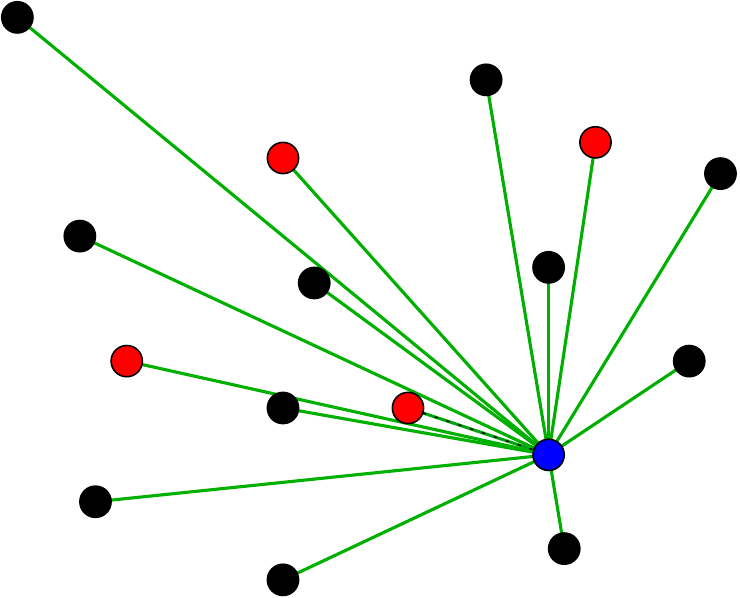}}\hfill\subfloat[Sampling]{\includegraphics[width=0.22\textwidth]{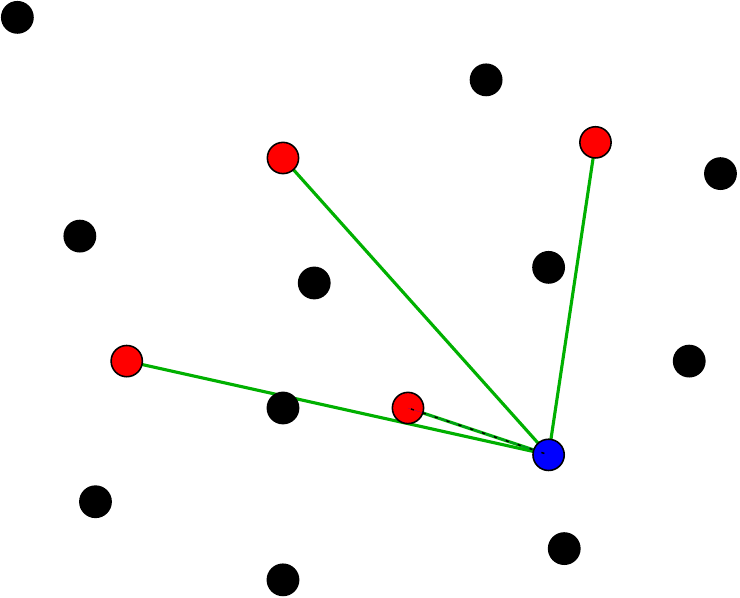}}\hfill\subfloat[Pivoting]{\includegraphics[width=0.22\textwidth]{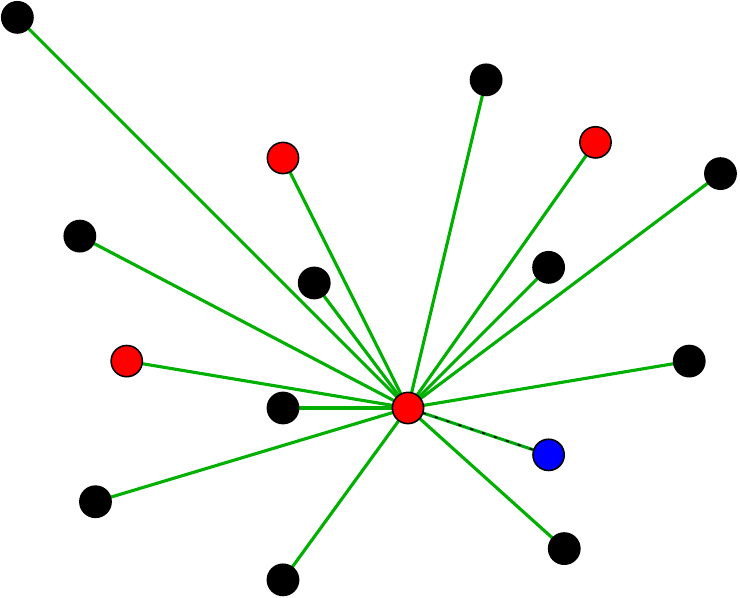}}\hfill\subfloat[Hybrid]{\includegraphics[width=0.22\textwidth]{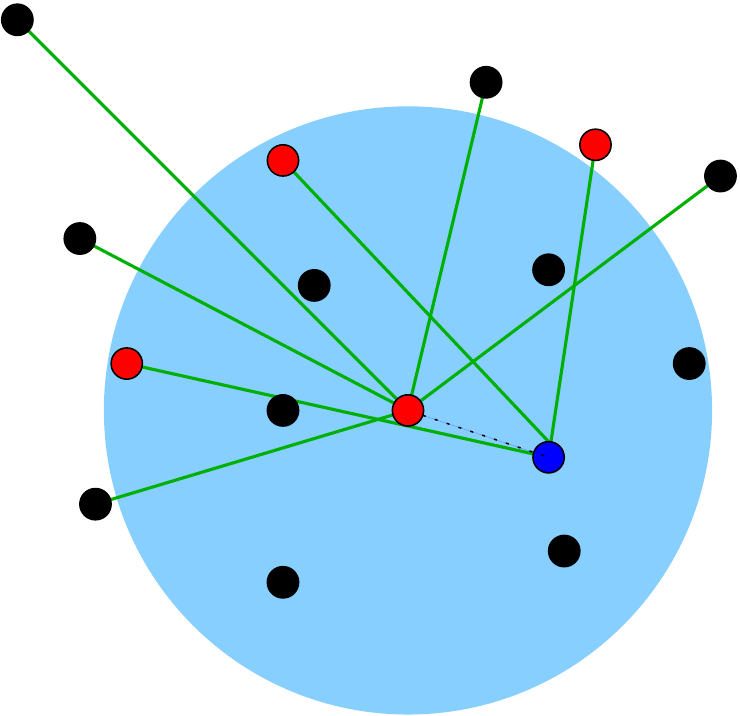}}\caption{Exact: Average distance from blue node to all other nodes. Sampling: Average distance to sampled (red) nodes. Pivoting: Average distance from pivot (closest sampled node). Hybrid: Distances outside the threshold radius from pivot are estimated through the pivot (but distances to sampled nodes outside the threshold are exact). Shorter distances, within the threshold radius, are estimated through sampled nodes. \label{epsh:fig}}\end{figure*}

Computationally, our algorithm selects a small uniform sample~$C$ of $k$ nodes and performs single-source shortest paths computation from each sampled node. We provide a high-level description, illustrated in Figure \ref{epsh:fig}, of how we use this information to estimate centralities of all nodes.

From the single-source computations, we know the distances from nodes in $C$ to all other nodes and therefore the exact value of $B(u)$ for each $u\in C$, but we need to estimate the centrality of other nodes. As we mentioned, a natural way to use this information is {\em sampling} \cite{Indyk:stoc1999,Thorup:icalp2001,EW_centrality:SODA2001,OkamotoCL:FAW2008}: Estimate the centrality of a node $v$ using the sample average $\hat{B}(v)=\sum_{u\in C} d_{vu}/k$. As we argued, however, the expected relative error can be very large when the distribution of distances from the node $v$ to all other nodes is skewed.

A second basic approach, which we propose here, is {\em pivoting}, which builds on techniques from approximate shortest-paths algorithms \cite{UY,Ecohen5f}. We define the {\em pivot} $c(v)\in C$ of a node $v$ as the node in the sample which is closest to $v$. We can then estimate the centrality of $v$ by that of its pivot, $B(c(v))$, which we computed exactly. By the triangle inequality, the value of~$B(v)$ is within~$\pm\,d_{v c(v)}$ of $B(c(v))$.

A large error, however, can be realized even on natural instances: The centrality of the center node in a star graph would be estimated with an error of almost $100\%$, using average distance of approximately 2 instead of 1. If we use the {\em pivoting upper bound}~$\hat{B}(v)=B(c(v))+d_{v c(v)}$ as our estimator, we obtain an estimate that is about three times the value of the true average. We can show, however, that this is just about the worst case: On all instances and nodes $v$, the pivoting upper bound estimate is, with high probability, not much less than $B(v)$ or much more than three times the value, that is, the estimate is within a factor of $3$ of the actual value. Since the argument is both simple and illuminating, we sketch it here. When the sample has size $k$, it is likely that the distance between $v$ and its pivot $c(v)$ is one of the~$1/k$ closest distances from $v$. Actually, with very high probability,~$d_{vc(v)}$ is one of the $(\log n)/k$ closest distances to $v$. Since $B(v)$ is the average value of a set of values such that~$(1-(\log n)/k)$ of them are at least as large as $d_{vc(v)}$, we obtain that \begin{equation}\label{arg1pivoting} B(v) \geq (1-(\log n)/k) d_{v c(v)}.\end{equation} We next apply the triangle inequality to obtain \begin{equation}\label{arg2pivoting} B(c(v))\leq B(v)+d_{v c(v)}. \end{equation} Finally, we combine~\eqref{arg1pivoting} and~\eqref{arg2pivoting} to obtain that our estimate~$\hat{B}(v) \equiv B(c(v))+d_{v c(v)} \leq B(v) +2 d_{v c(v)}$ is not likely to be much larger than $3 B(v)$.

Therefore, the pivoting estimator has a bounded error with high probability, regardless of the distribution of distances, a property we could not get with the sampling estimator. Neither method, sampling or pivoting, however, is satisfactory to us, since we are interested in a {\em small relative} error, for {\em all} nodes, on all instances, and with (probabilistic) guarantees.

Our key algorithmic insight is to carefully combine the sampling and pivoting approaches. When estimating centrality for a node $v$, we apply the pivoting estimate only to nodes $u$ that are ``far'' from $v$, that is, nodes that have distance $d_{vu}$ much larger than the distance to the pivot $c(v)$. The sampling approach is applied to the remaining ``closer'' nodes. By doing so, our hybrid approach obtains an estimate with a small relative error with high confidence, something that was not possible when using only one of the methods in isolation. Moreover, the computation needed by our hybrid algorithm is essentially the same as with the basic approaches: $k$ single-source shortest paths computation for a small value of $k$. Our hybrid estimator is presented and analyzed in Section~\ref{hybrid:sec}. The estimator is applicable to points in a general metric space and is therefore presented in this context. An efficient algorithm which computes the hybrid centrality estimate for all nodes in an undirected graphs is presented in Section~\ref{computeCC:sec}.

The effectiveness of our hybrid estimate in practice depends on setting a threshold correctly between pivoting and sampling. Our analysis sets a threshold with which we obtain guarantees with respect to worst-case instances, i.e., for any network structure and distances distribution of a node. In our implementation, we experiment with different settings. We also propose a novel {\em adaptive} approach, which estimates the error for several (or effectively all relevant) choices of threshold values, on a node per node basis. The sweet spot estimate which has the smallest estimated error is then used. Our error estimator for each threshold setting and our adaptive approach are detailed in Section~\ref{adaptive:sec}.

In applications, we are often interested in measuring centrality with respect to a particular topic or property which has a different presence at each node. Nodes can also intrinsically be heterogeneous, with different activity or importance levels. These situations are modeled by an assignment of weights $\beta(i) \geq 0$ to nodes. Accordingly, one can naturally define {\em weighted} classic closeness centrality of a node $i$ as \begin{equation} \label{Bclosenessbetacdef} B_\beta^{-1}(i) = \frac{\sum_{j\not=i } \beta(i)}{\sum_{j\not= i} \beta(i)d_{ij}}. \end{equation} In Section \ref{weighted:sec}, we present and analyze an extension of our algorithm designed for approximating weighted centralities. The approach is based on weighted sampling of nodes, which, for any weighting $\beta$, ensures a good approximations (small relative error) of Equation~\eqref{Bclosenessbetacdef}. The handling of weighted nodes is supported with almost no cost to scalability or accuracy when compared to unweighted instances.

In Section \ref{directed:Sec} we consider directed networks. When the graph is strongly connected, meaning that all nodes can reach all other nodes, it is often natural to consider closeness centrality with respect to {\em round-trip} distances. The round-trip distance between two nodes is defined as the sum $d_{uv}+d_{vu}$ of the shortest-paths distances. We show that a small modification of our hybrid algorithm, which requires both forward and reverse single-source shortest-paths computations from each sampled node, approximates round-trip centralities for all nodes with a small relative error. This follows because our hybrid estimator and its analysis apply in any metric space, and round-trip distances are a metric.

When the graph is not strongly connected, however, classic closeness centrality is not well defined: All nodes that have one or more unreachable nodes have centrality value of $0$. We may also want to separately consider inbound or outbound centralities, based on outbound distances from a node or inbound distances to a node, since these can be very different on directed graphs. Proposed modification of classic centrality to directed graphs are based on a combination of the average distance within the outbound or inbound reachability sets of a node, as well as on the cardinalities of these sets \cite{Linsocial:book,BoldiVigna:IM2014}. We therefore consider scalable estimation of these quantities, proposing a sampling-based solution which provides good estimates when the distance distribution is not too skewed.

Section~\ref{sec:related} briefly describes other relevant related work, including other important centrality measures. The results of our experimental evaluation are provided in Section \ref{experiments:sec}, demonstrating the scalability and accuracy of our algorithms on benchmark networks with up to tens of millions of nodes.

\section{The Hybrid Estimator} \label{hybrid:sec}

We present our hybrid centrality estimator, which applies for a set $V$ of $n=|V|$ points in a metric space.

We use parameters $k$ and $\epsilon$, whose setting determines a tradeoff between computation and approximation quality. We sample $k$ points uniformly at random from $V$ to obtain a set $C$. We then obtain the distances $d_{ij}$ from each point $i\in C$ to all points $j\in V$. The estimators we consider are applied to this set of $nk$ computed distances.

Specifically, we consider estimators $\hat{S}[j]$ for $j\in V$ of the sum~$S(j)=\sum_{i\in V} d_{ij}$. We then estimate the centrality of $j$ as~(the inverse of) $\hat{B}[j] \gets \hat{S}[j]/(n-1)$.

For points $j\in C$, we can compute the exact value of $S(j)$, since the exact distances $d_{ji}$ are available to all $i$. For $j\not\in C$ we are interested in estimating $S(j)$. We define the {\em pivot} of $j$ (closest node in the sample): $$c(j) = \arg\min_{i\in C} d_{ij}\ $$ and the distance $\Delta(j) = d_{j c(j)}$ to the pivot.

In the introduction we discussed three basic estimators: The {\em sample average} \begin{equation} \label{samp:est} \hat{B}(j) = \frac{1}{k} \sum_{i\in C} d_{ij}, \end{equation} the {\em pivot} estimator, $\hat{B}(j)\equiv B(c(j))$, and the {\em pivoting upper bound} $$\hat{B}(j)\equiv B(c(j))+\Delta(j).$$ We argued that neither one can provide a small relative error with high probability.

The hybrid estimate $\hat{S}[j]$ for a point $j\in V\setminus C$ is obtained as follows (efficient computation is discussed in the next section). We first compute the pivot $c(j)$ and its distance $\Delta(j)$. We then partition the points $V\setminus \{j\}$ to three parts $L(j)$, $\HC(j)$, and $H(j)$, where the placement of a node $i$ is determined according to its distance $d_{i c(j)}$ from the pivot $c(j)$. \begin{itemize} \item The points $L(j)$ ($L$ stands for ``low'') have distance at most~$\Delta(j)/\epsilon$ from $c(j)$. The sum of distances to these points is estimated using the sum of distances to the sampled points which are in $L(j)$. Since these points are a uniform sample from $L(j)$, we compute the {\em effective} sampling probability~$p(j) \equiv |L(j)\cap C|/|L(j)|$, and divide the sum by $p(j)$ to obtain an unbiased estimate. \item The set $\HC(j)$ (``high in $C$'') includes sampled points~$i\in C$ that have distance greater than $\Delta(j)/\epsilon$ from the pivot~$c(j)$. The distances from $v$ to these points are accounted for exactly. \item The set $H(j) \subset V\setminus C$ (``high'') are the points that are not sampled whose distance to the pivot $c(j)$ is greater than~$\Delta(j)/\epsilon$. The sum of distances to these points is estimated by the exact sum of their distances to $c(j)$. \end{itemize}

The estimate $\hat{S}[j]$ for $S(j)$ is thus \begin{equation} \label{estralg1:eq} \hat{S}[j]= \sum_{\mathclap{i\in H(j)}} d_{c(j) i} + \sum_{\mathclap{i \in \HC(j)}} d_{ji} + \frac{|L(j)|}{|L(j)\cap C|} \sum_{\mathrlap{i \in L(j)\cap C}} d_{ji}. \end{equation}

Since $c(j) \in L(j) \cap C$, the denominator satisfies $|L(j)\cap C|\geq 1$ and thus the estimator is well defined. It is easy to verify that the estimate $\hat{S}[j]$ for all points $j$ can be computed from the $nk$ distances we collected.

\subsection{Quality Guarantees}

We now analyse the quality of the hybrid estimator and show that the estimate $\hat{S}[j]$ has a small relative error for any point $j$:

\begin{theorem} \label{hybrid:thm} Using $k=1/\epsilon^3$, the hybrid estimator~\eqref{estralg1:eq} has a normalized root mean square error (NRMSE) of $O(\epsilon)$. Using~$k=\log n/\epsilon^3$, when applying the estimator to all points in $V$, we get a maximum relative error of $O(\epsilon)$ with high probability. \end{theorem} \begin{proof}

We consider the error we obtain by using $\hat{S}[j]$ instead of $S[j]$ for a point $j\in V\setminus C$. Error can be accumulated on accounting for distances to $H(j)$ or to $L(j)$.

The first set, $H(j)$, includes all non-sample points that have distance greater than $\Delta(j)/\epsilon$ from $c(j)$. The accumulated error on the sum is bounded by $\pm \Delta(j)$ for each point in $H(j)$. Since the distance from $j$ to a point in $i\in H(j)$ is at least $$d_{c(j)i}-\Delta(j)=\Delta(j)(1/\epsilon-1),$$ the relative error on all of $H(j)$ is at most $\Delta(j) / [\Delta(j) (1/\epsilon-1)] = 1 / (1/\epsilon-1) = \epsilon/(1-\epsilon)$.

We now turn to $L(j)$, where we use a sampling estimator: We estimate the sum of distances to points in $L(j)$ using the sum of distances to sample points that are in $L(j)$. The sample points constitute a random sample of $L(j)$, which includes each point in~$L(j)$ with probability $p=k/n$.

We compute the variance of estimating $\sum_{i\in L(j)} d_{ji}$ using the estimate $\frac{1}{p}\sum_{i\in L(j) \cap C} d_{ji}$. Consider the ratio of the variance to the square of the sum. The ratio is maximized when the set $L(j)$ includes all points (otherwise the contribution of $H(j)$ increases the denominator but not the numerator). Therefore, since we are upper bounding the error, we can assume that the set $L(j)$ contains all points.

The points in $L(j)$ are of distance at most $\Delta(j)(1/\epsilon+1)$ from~$j$.

We first consider the total contribution to the centrality of the set of points $A$ that are of distance smaller than $\Delta(j)$ from $j$. Since $\Delta(j)$ is the distance to the pivot, the expected number of such points is not more than $n/k$. Their expected total relative contribution to $B(j)$ is at most their relative fraction, which in expectation is $1/k \ll \epsilon$. Moreover, for an integer $a>1$, the probability of there being more than $an/k$ such points is the probability that all $k$ sampled points selected among the $n(1-a/k)$ farthest points from $j$, which is at most $(1-a/k)^k \approx e^{-a}$. So the contribution of points $A$ to centrality (and to the variance) is also well concentrated.

We now consider the contribution to variance of points that have distance between $\Delta(j)$ and $(1/\epsilon+1) \Delta(j)$. For convenience we use $s\equiv 1/\epsilon+1$ and $\Delta \equiv \Delta(j)$. Repeating the same argument as before, since we are computing an upper bound we can assume that this set contains all points. Given the sum of distances of these points, the ``worst case'' for variance is when all distances are at one of the extremes; we thus further assume that the distance of each point is either $\Delta$ or $s \Delta$. The variance contribution of a point is $(1/p-1)$ times its distance squared. We now define~$x\in [0,1]$ to be the fraction of points are of distance $\Delta$; the remaining have distance $s \Delta$. The sum of distances is $$n (x\Delta + (1-x) s \Delta)=n\Delta(x+(1-x)s)$$ and the variance is \begin{align*} & ((1/p) -1) n (x \Delta^2 + (1-x)s^2 \Delta^2) \\ & = ((1/p) -1)n \Delta^2 (x+ (1-x)s^2) \\ & \leq \frac{n^2 \Delta^2}{k}(x+ (1-x)s^2). \end{align*} We now consider the maximum over choices of $n$ and $x$ of the ratio of the variance to the square of the mean, which is $$\max_{x\in [0,1]} \frac{1}{k} \frac{x + (1-x)s^2}{(x+(1-x)s)^2}.$$ This is maximized at $x=s/(s+1)=(1+\epsilon)/(1+2\epsilon)$. The maximum is $\frac{1}{k}\frac{(s+1)^2}{4s}=\frac{(1+2\epsilon)^2}{4k\epsilon (1+\epsilon)} \approx \frac{1}{4k\epsilon}$. This means that the Coefficient of Variation (CV) is about $\frac{1}{2\sqrt{k\epsilon}}$.

Balancing the sampling CV with the pivoting relative error of $\epsilon$ we obtain $k \approx \frac{1}{2\epsilon^3}$. \end{proof}

In our implementation, we worked with parameter settings of~$\epsilon=\sqrt{k}$. This setting means that the relative error on the pivoting component is at most $\epsilon/(1-\epsilon)$. We can typically expect it to be much smaller, however. First, because distances in $H(j)$ can be much larger than $\Delta(j)/\epsilon$. Second, the estimates of different points are typically not ``one sided'' (the estimate is one sided when the pivot happens to be on or close to the shortest path from $j$ to most other points), so errors can cancel out. For the sampling component, the analysis was with respect to a worst-case distance distribution, where all values lie at the extremes of the range, but in practice we can expect an error of $\approx 1/\sqrt{k} \approx \epsilon$. Moreover, when the population variance of $L(j)$ is small, we can expect a smaller relative error.

In Section \ref{adaptive:sec} we propose adaptive error estimation, which for each point $j$, uses the sampled distances $d_{ij}$ to obtain a tighter estimate on the actual error.

\section{Computing Estimates} \label{computeCC:sec}

We now consider closeness centrality on undirected graphs, with a focus on efficient computation, both in terms of running time and the (run-time) storage we use, Specifically, we would like to compute estimates $\hat{S}[v]$ of $S(v)=\sum_{j} d_{vj}$ for all nodes $v\in V$.

All the estimators we consider, the basic sampling \eqref{samp:est} and pivoting estimates and the hybrid estimate \eqref{estralg1:eq} are applied to a set of (at most) $kn$ sampled distances. To compute these distances, we can first sample a set $C$ of $k$ nodes uniformly at random and then run Dijkstra's single-source shortest path algorithm from each node $u\in C$ to compute the distances $d_{uv}$ from $u$ to all other nodes. The computation of the estimates $\hat{S}[v]$ given these distances is linear. The issue with this approach is a run-time storage of $O(nk)$.

We first observe that both the basic sampling and the basic pivoting estimates can be computed using only $O(1)$ run-time storage per node. With sampling, we accumulate, for each node~$v$, the sum of distances from the nodes in $C$. We initialize the sum to $0$ for all $v$ and then when running Dijkstra from $u\in C$, we add $d_{uv}$ to each scanned node $v$. The additional run-time storage used here is the state of Dijkstra and $O(1)$ additional storage per node. With pivoting, we initialize $\Delta(v)\gets \infty$ for all nodes. When running Dijkstra from $u$, we accumulate the sum of distances as~$S(v)$. We also update $\Delta(v) \gets \min\{d_{uv},\Delta(v)\}$ when a node~$v$ is scanned. When $\Delta(v)$ is updated, we also update the pivot~$c(v) \gets u$. Finally, for each node $v$, we estimate $S(v)$ by the precomputed $S(c(v))$.

The pseudocode provided as Algorithm \ref{bavelasu:alg} computes the hybrid estimates \eqref{estralg1:eq} for all nodes using $O(1)$ additional storage per node. To do so with only $O(1)$ storage, we use an additional run of Dijkstra: For each node $v\in V$, we first compute its pivot $c(v)$ and the distance $\Delta(v) = d_{v c(v)}$. This can be done with a single run of Dijkstra's algorithm having all sampled nodes as sources.

We then run Dijkstra's algorithm from each sampled node~$u\in C$. For the sampled nodes $u\in C$, the sum $S(u)$ is computed exactly; for such cases, we have $\hat{S}[u]=S(u)$. For the nodes $v\not\in C$ we compute an estimate $\hat{S}[v]$.

The computation of the estimate is based on identifying the three components of the partition of $V\setminus \{v\}$ into $L(v) \cup \HC(v) \cup H(v)$, which is determined according to distances from the pivot~$c(v)$. The pivot mapping computed in the additional run is used to determine this classification.

The contributions to the sum estimates $\hat{S}[v]$ are computed during the single-source shortest paths computations from $C$. In particular, the contribution to $\hat{S}[v]$ of sampled nodes $u\in L(v)\cup \HC(v)$ are computed when we run Dijkstra from $u$. The contribution of $H(v)$ is computed when we run Dijkstra from the pivot $c(v)$ of $v$.

When running Dijkstra from a sampled node $u\in C$ and visiting~$v$, we need to determine whether $u$ is in $L(v)$ or $\HC(v)$ in order to compute its contribution. If $u\in \HC(v)$, we increase $\hat{S}[v]$ by $d_{uv}$. If $u\in L(v)$, we would like to increase $\hat{S}[v]$ by $d_{uv}/p[v]$. At that point, however, $p[v]$, which depends on $|L(v)|$ and $|C\cap L(v)|$, may not be available. We therefore add $d_{uv}$ to $\text{\sc LCsum}[v]$, which tracks the sum of distances to nodes in $C\cap L(v)$. We also increment $\text{\sc LCnum}[v]$, which tracks the cardinality $|C\cap L(v)|$. When the $k$ Dijkstra runs terminate, we can compute $p[v]$ and increase~$\hat{S}[v]$ by $\text{\sc LCsum}[v]/p[v]$.

Deciding whether $u$ is in $L(v)$ or $\HC(v)$ can sometimes be done only after the pivot $c(v)$ was visited by the Dijkstra run from $u$. If $d_{uv}> \Delta(v)(1/\epsilon+1)$ then from the triangle inequality~$d_{uc(v)}> \Delta(v)/\epsilon$ and we can determine that $u\in \HC(v)$. Similarly, if $d_{uv}\leq \Delta(v)(1/\epsilon-1)$ we can determine that $u\in L(v)$. Otherwise, we can classify $u$ only after we visit $c(v)$ and know the distance~$d_{c(v) u}$. In this case, the accounting of $u$ to $\hat{S}[v]$ is postponed: We place the pair $(v,d_{uv})$ in {\sc List}$[c(v)]$. Each time a sampled node $z\in C$ is visited by $u$, we process the list {\sc List}$[z]$ and for each entry~$(v,d_{vu})$ we use $d_{uz}\equiv d_{uc(v)}$ to classify $v$ and accordingly increase $\hat{S}[v]$ or~$\text{\sc LCsum}[v]$. {\sc List}$[z]$ is then deleted.

The accounting for $H(v)$ is done when running Dijkstra from the pivot $c(v)$. During Dijkstra from $u$, we record information on each node $v$ for which $c(v)\equiv u$. The threshold values $\Delta(v)/\epsilon$ are recorded in increasing order in the {\sc Thresh} array, as nodes are visited. The set of nodes with pivot $u$ and a threshold value is recorded in the entry of {\sc Nodes} which corresponds to the threshold value. The sum of distances from $u$ to all nodes in~$V\setminus C$ with distances that are between entries in the {\sc Thresh} array is computed in the corresponding entries of the {\sc Bin} array. After Dijkstra's algorithm from $u$ is completed, we process these arrays in reverse, computing for each node $v$ such that $c(v)\equiv u$ the contribution of $H(v)$ to the estimate $\hat{S}[v]$.

This algorithm performs $k+1$ runs of Dijkstra's algorithm and uses running storage that is linear in the number of nodes~(does not depend on $k$). This means the algorithm has very little computation overhead over the basic estimators.

\begin{algorithm*}\caption{Centrality estimation for all nodes: undirected\label{bavelasu:alg}} \begin{algorithmic} \State {\bf Input:} Network $G$, integer $k> 0$, $\epsilon>0$ \State select uniformly at random $k$ nodes $C=\{c_1,\ldots,c_k\}\subset V$ \For {$v\in V$} \Comment{Computation equivalent to a single Dijkstra} \State $c[v]\gets \arg\min_{i=1,\ldots,k} d_{c_i v}$ \Comment{Pivot of $v$} \State $\Delta[v] \gets d_{v,c_{c[v]}}$ \Comment{distance of $v$ to its pivot} \State $\hat{S}[v]\gets 0$ ; $\text{\sc LCsum}[v] \gets 0$; $\text{\sc LCnum}[v] \gets 0$ ; $\text{\sc LCsumSq}[u] \gets 0$ ; $\text{\sc HCsum}[u] \gets 0$ ; $\text{\sc HCsumSqErr}[u] \gets 0$ ; \EndFor \For {$i=1,\ldots,k$} \State $t\gets 0$; $curt\gets 0$; {\sc Thresh}$[0]\gets 0$ \Comment{Initialize thresholds array and counters} \State Run Dijkstra from the sampled node $c_i$ \For {each new node $u$ visited by Dijkstra} \State $d\gets d_{c_i u}$ \Comment{distance from $c_i$ to $u$} \State $\hat{S}[c_i]\gets \hat{S}[c_i]+d$ \If {$u \in C$} \Comment{equivalently, $c_{c[u]} = u$} \State $j\gets c[u]$ \Comment{a sampled node is its own pivot, we get its index} \State $\text{\sc last}[j] \gets i$; $\text{\sc dist}[j] \gets d$ \Comment{$c_j$ was visited from $c_i$ and has distance $\text{\sc dist}[j]$} \For {$z \in \text{\sc List}[j]$} \If {$d > \Delta[z.node]/\epsilon$} $\text{\sc HCsum}[z.node] \overset{+}{\gets} z.d$ \Comment{$c_i\in HC(z.node)$} \State $\text{\sc HCsumSqErr}[z.node] \overset{+}{\gets} (z.d-d)^2$ \Else $\text{\ \sc LCsum}[z.node] \overset{+}{\gets} z.d$; $\text{\sc LCnum}[z.node] \overset{+}{\gets} 1$; $\text{\sc LCsumSq}[z.node] \overset{+}{\gets} z.d^2$ \Comment{$c_i\in L(z.node)$} \EndIf \EndFor \State Delete $\text{\sc List}[j]$ \Else \Comment{$u\not\in C$} \If {($d \leq \Delta[u](1/\epsilon-1)$) {\bf or} ($\text{\sc last}[c[u]] = i$) {\bf and} ($\text{\sc dist}[c[u]] \leq \Delta[u]/\epsilon$) } \Comment{$c_i \in L(u)$} \State $\text{\sc LCsum}[u] \overset{+}{\gets} d$; $\text{\sc LCnum}[u] \overset{+}{\gets} 1$ \State $\text{\sc LCsumSq}[u] \overset{+}{\gets} d^2 $ \Else \Comment{We can not determine if $c_i\in L(u)$ or we know $c_i\in HC(u)$ but $c[u]$ was not yet visited} \State $z.node \gets u$; $z.d \gets d$ \State $\text{\sc List}[c[u]] \gets \text{\sc List}[c[u]] \cup \{z\}$ \EndIf

\If {$c[u] = i$} \Comment{$c_i$ is the pivot of $u$} \If {\text{\sc Thresh}$[t] = d/\epsilon$} \Comment{same threshold as previous} \State $\text{\sc nodes}[t]\gets \text{\sc nodes}[t] \cup \{ u\} $ \Else $\ t\gets t+1$; $\text{\sc Thresh}[t] \gets d/\epsilon$; $\text{\sc nodes}[t]\gets \{ u\}$; $\text{\sc bin}[t] \gets 0$; $\text{\sc count}[t]\gets0$ \EndIf \EndIf

\While {$curt<t$ \textbf{and} $d > \text{\sc Thresh}[curt+1]$} \increase{curt}{1} \EndWhile \If {$d > \text{\sc Thresh}[curt]$} \increase{\text{\sc bin}[curt]}{d}; \increase{\text{\sc count}[curt]}{1} \EndIf \EndIf

\EndFor

\Comment{Compute tail sums for nodes for which $c_i$ is pivot} \State $\text{\sc tailsum}\gets 0$; $\text{\sc tailnum}\gets 0$ \While {$t>0$} \State $\text{\sc tailsum} \overset{+}{\gets} \text{\sc bin}[t]$ \State $\text{\sc tailnum} \overset{+}{\gets} \text{\sc count}[t]$ \For {$u \in \text{\sc nodes}[t]$} \State $\text{\sc Hsum}[u] \gets \text{\sc tailsum}$ \State $\text{\sc Hnum}[u] \gets \text{\sc tailnum}$ \Comment{ {\sc Hnum}$[u] = |H(u)|$; {\sc Hsum}$[u] = \sum_{v\in H(u)} d_{c(u) v}$ } \EndFor \State $t\gets t-1$ \EndWhile \EndFor \For {$u\in V \setminus C$} \State $\text{\sc Lnum} \gets n-1-\text{\sc Hnum}[u] -k + \text{\sc LCnum}[u]$; $\text{\sc HCnum} \gets k-\text{\sc LCnum}$ \State $p \gets \frac{\text{\sc LCnum}[u]}{\text{\sc Lnum}}$ \Comment{Fraction of sampled nodes that are in $L(u)$} \State $\hat{S}[u] \gets \text{\sc Hsum}[u] +\text{\sc HCsum}[u] + \text{\sc LCsum}[u]/p$ \State $\text{\sc SqErrEst}[u] \gets \frac{1}{\text{\sc LCnum}[u]}(\frac{\text{\sc LCsumSq}[u]}{\text{\sc LCnum}[u]}- \big(\frac{\text{\sc LCsum}[u]}{\text{\sc LCnum}[u]}\big)^2) \text{\sc Lnum}[u] + \frac{\text{\sc HCsumSqErr}[u]}{\text{\sc HCnum}}\text{\sc Hnum}[u]$ \EndFor \Return {For all $u$: ($u,\hat{S}[u],\text{\sc SqErrEst}[u]$)} \end{algorithmic} \end{algorithm*}

\section{Adaptive Error Estimation} \label{adaptive:sec}

Algorithm~\ref{bavelasu:alg} also computes, for each node $v$, an estimate on the error of our estimate $\hat{S}[v]$. This estimate is {\em adaptive}, that is, it depends on the input. This is in contrast to the error bounds in Theorem \ref{hybrid:thm}, which are with respect to {\em worst-case} instances and, if used, will typically grossly overestimate the actual error and provide weak and pessimistic confidence bounds. We explain how these adaptive estimates are computed.

We also propose {\em adaptive error minimization} as Algorithm \ref{bavelaserr:alg}: Instead of working with a fixed value of $\epsilon$, as in Algorithm \ref{bavelasu:alg}, the new algorithm chooses the estimate that has the smallest estimated error.

\subsection{Error Estimation}

In Algorithm~\ref{bavelasu:alg}, error estimates are computed separately for each of the two components: one from the pivoting on the ``distant'' nodes $H(v)$, and one from the sampling, on the ``closer'' nodes~$L(v)$.

The pivoting error is estimated by considering distant sampled nodes, that is, nodes in $\HC(v)$. These nodes are treated as a representative sample of $H(v)$. For these nodes, we take the average of the squared difference between the distance of the node from $v$ and its distance from the pivot $c(v)$: \begin{equation} \label{higherr} \widehat{SQ}(H(v))=\frac{1}{|\HC(v)|} \sum_{\mathrlap{u\in \HC(v)}} \big(d_{uv}-d_{c(v) u} \big)^2. \end{equation} Note that for nodes in $\HC(v)$, both these distances are available from the single-source shortest-paths computations we performed. Finally, to obtain an estimate on the contribution of the pivoting component to the squared error of $\hat{S}[v]$, we multiply by the magnitude $|H(v)|$ of the set $H(v)$, which we know exactly. In cases when there are not enough or no samples (when $\HC(v)$ is empty), we instead compute the average squared difference over a ``suffix'' of the farthest nodes in $C$.

The sampling error applies to the remaining ``closer'' nodes~$L(v)$ and depends on the distribution of distances in $L(v)$, that is, on the population variance of $L(v)$, and on the sample size from this group, which is $L(v)\cap C$. We first estimate the population variance of the set of distances from $v$ to the set of nodes $L(v)$. This is estimated using the sample variance of the uniform sample~$L(v)\cap C$, as \begin{align} \hat{\sigma^2}(L(v)) &= \frac{1}{|C\cap L(v)|} \sum_{u\in C\cap L(v)} \left(d_{uv} - \frac{\sum_{u\in C\cap L(v)} d_{uv}}{|C\cap L(v)|}\right)^2 \nonumber\\ &= \frac{\sum_{u\in C\cap L(v)} d_{uv}^2}{|C\cap L(v)|}-\left(\frac{\sum_{u\in C\cap L(v)} d_{uv}}{|C\cap L(v)|}\right)^2. \label{lowerr} \end{align} We then divide the estimated population variance by the number of samples $|L(v)\cap C|$ (variable {\sc LCnum} in the pseudocode) to estimate the variance of the average of $|L(v)\cap C|$ samples from the population. To estimate the variance contribution of the sampling component to the sum estimate $\hat{S}[v]$, we multiply by~$|L(v)|$~(variable {\sc Lnum} in the pseudocode). The combined square error of~$\hat{S}[v]$ is estimated by summing these two components: $$|H(v)| \widehat{SQ}(H(v)) + \frac{|L(v)|}{|L(v)\cap C|} \hat{\sigma^2}(L(v)).$$

\subsection{Adaptive Error Minimization}

In order to get the most mileage from the $k$ single source shortest paths computations we performed, we would like to adaptively select the best ``threshold'' between pivoting and sampling, rather than work with a fixed value.

For a node $v\in V$ and a threshold value $T$ let \begin{align*} H(v,T) &= \{ u\in V\setminus C \mid d_{c(v) u}> T \} \\ \HC(v,T) &= \{ u\in C \mid d_{c(v) u}> T \} \\ L(v,T) &= \{ u\in V \mid d_{c(v) u} \leq T \}. \end{align*} The set $H(v,T)$ contains all non-sampled nodes with distance from $c(v)$ greater than $T$, the set $\HC(v,T)$ contains all sampled nodes with distance from $c(v)$ greater than $T$, and the set $L(v,T)$ contains all nodes with distance from $c(v)$ at most $T$.

We can then define an estimator with respect to a threshold $T$, as in Equation \eqref{estralg1:eq}: \begin{equation}\label{eqn:thresholdestimate} \hat{S}(v,T) = \sum_{\mathrlap{u\in H(v,T)}} d_{c(v) u} + \sum_{\mathrlap{u \in \HC(v,T)}} d_{vu} + \frac{|L(v,T)|}{|L(v,T)\cap C|} \sum_{\mathrlap{u \in L(v,T)\cap C}} d_{vu}. \end{equation}

In Algorithm \ref{bavelasu:alg} we used the threshold value $T_v = \Delta(v)/\epsilon$ for a node $v$. Here we choose $T_v$ adaptively so as to balance the estimated error of the first and third summands.

One way to achieve this is to apply Algorithm \ref{bavelasu:alg} simultaneously with several choices of $\epsilon$. Then, for each node, we take the value with the smallest estimated error. We propose here Algorithm \ref{bavelaserr:alg}, which maintains $O(k)$ state per node but looks for the threshold sweet spot while covering the full range between pure pivoting and pure sampling.

Algorithm \ref{bavelaserr:alg} computes estimates and corresponding error estimates as in Algorithm \ref{bavelasu:alg}. The estimates, however, are computed for $k$ values of the threshold $T_v$ which correspond to the distances from $c(v)$ to each of the other sampled nodes. From these $k$ estimates, the algorithm selects the one which minimizes the estimated error.

 The reason for considering only these $k$ threshold values (for each pivot) is that they represent all the possible assignments of sampled nodes to $L(v)$ or $\HC(v)$.

Finally, we note that the run-time storage we use depends linearly in the sets of threshold values and therefore it can be advantageous, when run-time storage is constrained, to reduce the size further. One way to do this is, for example, to only use values of $T_v$ which correspond to discretized distances.

 \begin{algorithm*}\caption{Classic closeness centralities with adaptive error minimization\label{bavelaserr:alg}} \begin{algorithmic} \State select a set $C=\{c_1,\ldots,c_k\}\subset V$ of sampled nodes, uniformly at random; for $j=1,\ldots,k$, use $c[c_j] \gets j$. \For {$v\in V$} $\Delta[v]\gets \infty$ \EndFor \For {$i=1,\ldots,k$} \State $\Delta[c_i]\gets 0$ \Comment{pivot of $c_i$ is itself, distance to pivot is $0$} \State $cvisited\gets 1$; $vvisited \gets 0$ \Comment{number of nodes in $C$ and $V\setminus C$, respectively, visited so far} \State $distsumvisited \gets 0$ \Comment{sum of distances to nodes in $V\setminus C$ visited so far} \State $\delta[i,i]\gets 0$ \Comment{$\delta[i,j]$ is the distance between sampled nodes $c_i$ and $c_j$} \State $\pi[i,1]\gets i$ \Comment{$\pi[i,*]$ is the permutation of sampled nodes by increasing distance from $c_i$} \State Run Dijkstra's algorithm from $c_i$ \For {$v\in V$ in order of first visit by Dijkstra} \State $d\gets d_{c_i v}$ \If {$v \in C$} \State $j \gets c[v]$ \Comment{index of sampled node $v$} \State $cvisited\gets cvisited+1$; $\pi[i,cvisited]\gets j$; $\delta[i,j] \gets d$ \State $\text{\sc TailNum}[i,cvisited] \gets vvisited$ \State $\text{\sc TailSum}[i,cvisited] \gets distsumvisited$ \Else \Comment{$v\not\in C$} \If {$d < \Delta[v]$} \State $\Delta[v] \gets d$, $c[v]\gets i$ \EndIf \State $D[v,i] \gets d$ \Comment{$(n-k)\times k$ matrix of distances of $v\in V\setminus C$ to sampled nodes $1,\ldots,k$} \State $vvisited\overset{+}{\gets}1$; $distsumvisited \overset{+}{\gets} d$ \EndIf \EndFor \State After Dijkstra ends: \For {$j=1,\ldots,k$} \State $\text{\sc TailNum}[j,cvisited] \gets vvisited-\text{\sc TailNum}[j,cvisited]$ \State $\text{\sc TailSum}[j,cvisited] \gets distsumvisited-\text{\sc TailSum}[j,cvisited]$ \EndFor \State $\hat{S}[c_i] \gets distsumvisited + \sum_{j=1}^k \delta[i,j]$ \Comment{Exact $S[c_i]$ of sampled node $c_i$} \State $\text{\sc EstErr}[c_i]\gets 0$; \Comment{estimated errors (no errors) for $\hat{S}[c_i]$. } \EndFor \For {$v\in V\setminus C$} \Comment{Compute $\hat{S}$, $\text{\sc EstErr}$ for all remaining nodes} \State $\text{\sc LCsum} \gets 0$; $\text{\sc HCsum} \gets \sum_{i=1}^k D[v,i]$; $\text{\sc HCsumSqErr} \gets \sum_{i=1}^k (D[v,i] - \delta[c(v),i])^2 $ \State $\hat{S}[v] \gets \hat{S}[c[v]]$; $\text{\sc EstErr}[v] \gets \text{\sc HCsumSqErr} \cdot (n-1-k)/k$ \State $\text{\it MinErr} \gets \text{\sc EstErr}[v]$ \For {$i=1,\ldots,k$} \Comment{scan sampled nodes $\pi[c(v),i]$ by increasing distances from $c(v)$} \State $\text{\sc LCsumSq} \overset{+}{\gets} D[v,\pi[c(v),i]]^2$ \State $\text{\sc Hnum} \gets \text{\sc TailNum}[c(v),\pi[c(v),i]]$ \State $\text{\sc Lnum} \gets n-1-\text{\sc Hnum} -k+i$ \Comment{$|L(v)|$ for current threshold} \State $\text{\sc LCnum} \gets i$; $p \gets \text{\sc LCnum}/\text{\sc Lnum}$ \State $\text{\sc LCsum} \overset{+}{\gets} D[v,\pi[c(v),i]]$ \Comment{sum of distances to sampled nodes within threshold} \State $\text{\sc HCsum} \overset{-}{\gets} D[v,\pi[c(v),i]]$ \Comment{sum of distances to sampled nodes outside threshold} \State $\text{\sc Hsum} \gets \text{\sc TailSum}[c(v),\pi[c(v),i]]$ \State $\text{\sc HCsumSqErr} \overset{-}{\gets} (D[v,\pi[c(v),i]]-\delta[c(v),\pi[c(v),i]])^2$ \State $\text{\sc Est} \gets \text{\sc LCsum}/p + \text{\sc Hsum} + \text{\sc HCsum}$ \Comment{estimated $S[v]$} \State $\text{\sc EstErr} \gets \frac{1}{\text{\sc LCnum}}(\frac{\text{\sc LCsumSq}}{\text{\sc LCnum}}- \big(\frac{\text{\sc LCsum}}{\text{\sc LCnum}}\big)^2) \text{\sc Lnum} + \frac{\text{\sc HCsumSqErr} }{\text{\sc HCnum}} \text{\sc Hnum}$\Comment{est.\ error for threshold $\delta[c(v),\pi[c(v),i]]$} \If {$\text{\sc EstErr} < \text{\it MinErr}$} \State $\text{\it MinErr} \gets \text{\sc EstErr}$ \Comment{Look for the estimation sweet spot} \State $\hat{S}[v] \gets \text{\sc Est}$; $\text{\sc SqErrEst}[v] \gets \text{\sc EstErr}$ \EndIf \EndFor \EndFor \Return {$\hat{S},\text{\sc SqErrEst}$} \end{algorithmic} \end{algorithm*}

\section{Weighted Centrality} \label{weighted:sec}

We now consider weighted classic closeness centrality with respect to node weights $\beta:V \geq 0$, as defined in Equation~\eqref{Bclosenessbetacdef}. We limit our attention to estimating the denominator $$S_\beta(i) = \sum_{j\not= i} \beta(i)d_{ij},$$ since the numerator $\sum_{j\not=i } \beta(i)$ can be efficiently computed exactly for all nodes by computing the sum $\sum_i \beta(i)$ once and, for each node $j$, subtracting the weight of the node $j$ itself from the total. We show how to modify Algorithm \ref{bavelasu:alg} to compute estimates for $S_\beta(i)$ for all nodes. We will also argue that the proof of Theorem~\ref{hybrid:thm} goes through with minor modifications, that is, we obtain a small relative error with high probability.

If the node weights are in $\{0,1\}$, the modification is straightforward. We obtain our sample $C$ only from nodes $i$ with weight~$\beta(i)=1$ and account only for these nodes in our estimate of $S$.

We now provide details on the modification needed to handle general weights $\beta$. The first component is the node sampling. We apply a weighted sampling algorithm; in particular, we use~$\text{\sc VarOpt}$ stream sampling \cite{varopt_full:CDKLT10,Cha82}, which is a weighted version of reservoir sampling \cite{Knuth2f,Vit85}. We obtain a sample of exactly $k$ nodes so that the inclusion probability of each node is proportional to its weight. More precisely, $\text{\sc VarOpt}$ computes a threshold value~$\tau$ (which depends on $k$ and on the distribution of~$\beta$ values). A node $v$ is sampled with probability $\min\{1,\beta(v)/\tau\}$. These sampling probabilities are PPS (Probability Proportional to Size), but with $\text{\sc VarOpt}$ we obtain a sample of size exactly~$k$ (whereas independent PPS only guarantees an expected size of $k$). For each sampled node we define its {\em adjusted weight} $\hat{\beta}(v) = \max\{\tau,\beta(v)\}$, where $\tau$ is the {\sc VarOpt} threshold.

The weighted algorithm is very similar to Algorithm~\ref{bavelasu:alg}, but requires the modification stated as Algorithm \ref{weighted:alg}. The contributions to $\hat{S}[u]$ of nodes $v$ that are in $H[u]$ (accounted for in the tail sums computed in the $\text{\sc bin}$ array) or in $\HC[u]$ are multiplied by $\beta(v)$. For nodes in $L(v)$, we compute the inverse probability estimate with respect to the inclusion probability $\min\{1,\beta(v)/\tau\}$. We divide the contribution, which is $\beta(v)d_{uv}$, by the inclusion probability, obtaining $\hat{\beta}(v)d_{uv}$.

Our error estimates can also be easily modified to work with weighted centralities. Instead of the cardinality of each set, we use the total $\beta$ weight of the set; instead of a sum of distances, we use the $\beta$-weighted sum.

{\small \begin{algorithm} \caption{Modifications of Alg. \ref{bavelasu:alg} for weighted centrality\label{weighted:alg}} \begin{algorithmic} \State $\hat{S}[c_i] \overset{+}{\gets}\beta(u)d_{c_i u}$ \Comment{when computing $\hat{S}$ for $c_i\in C$} \State $\hat{S}[u] \overset{+}{\gets} \beta(c_i) d_{c_i u}$ \Comment{when $c_i\in HC(u)$} \State $\hat{S}[u] \overset{+}{\gets} \hat{\beta}(c_i) d_{c_i u}$ \Comment{when $c_i\in L(u)$} \If {$\beta(c_i) < \tau$} $\text{\sc varest}[u] \overset{+}{\gets} d_{c_i u}^2 (\tau-\beta(c_i))\tau$ \Comment{when $c_i\in L(u)$; when $\beta(c_i)> \tau$ then $c_i$ is included with probability $1$ and its contribution to variance is $0$.} \EndIf

\State $\text{\sc bin}[curt] \overset{+}{\gets} \beta(u) d_{c_i u}$ ; $\text{\sc count}[curt] \overset{+}{\gets} \beta(u)$ \Comment{when computing tail sums/counts for $c_i$} \end{algorithmic} \end{algorithm} }

The analysis of the approximation quality of $\hat{S}$ in Algorithm~\ref{weighted:alg} carries over to the weighted algorithm. In fact, the skewness of~$\beta$ can only improve estimation quality: intuitively, the sample would contain in expectation more than $k/n$ fraction of the total~$\beta$ weight, since heavier items are more likely to be sampled.

\section{Directed graphs} \label{directed:Sec}

\subsection{Round-trip Centralities} For a strongly connected directed graph, it is natural to consider the round-trip distances $\overleftrightarrow{d}_{ij} \equiv d_{ij}+d_{ji}$, and {\em round-trip centrality} values computed with respect to these round-trip distances.

Since round-trip distances are a metric, the hybrid estimator~\eqref{estralg1:eq} applies, as does Theorem \ref{hybrid:thm}, which provides the strong guarantees on approximation quality. Moreover, a simple modification of the algorithms we presented for undirected graphs applies to estimation of round-trip centralities in strongly connected directed graphs. We choose a uniform random sample of $k$ nodes, as we did in the undirected case. Then, for each sampled node $u\in C$, we perform two single-source shortest paths computations, to compute the forward and a backward distances to all other nodes. Then for each node $v\in V\setminus C$, we compute the sum $\overleftrightarrow{d}_{uv} = d_{uv}+d_{vu}$ of these distances. We sort the nodes $v$ by increasing $\overleftrightarrow{d}_{uv}$. We then use the sorted order and round-trip distances the same way we used the Dijkstra order in the undirected version of the algorithm.

\subsection{Inbound and Outbound Centralities} As mentioned in the introduction, for general (not necessarily strongly connected) directed graphs, we may also be interested in separating {\em outbound} or {\em inbound centralities}. In particular, we are interested in the average distance from a particular node $v$ to all nodes it can reach (outbound centrality) or from nodes that can reach $v$ (inbound centrality), as well as in the cardinalities of these sets.

The size of the outbound reachability set of $v$ is $$\overrightarrow{R}[v] = \left| \{u\in V\setminus\{v\} \mid v \leadsto u \} \right|,$$ where $v \leadsto u$ indicates that $u$ is reachable from $v$. Similarly, the size of the inbound reachability set of $v$ is $$\overleftarrow{R}[v] = \left| \{u\in V\setminus\{v\} \mid u \leadsto v \} \right|.$$ Accordingly, we define the total distance to the outbound reachability set of $v$ as $$\overrightarrow{S}[v] = \sum_{\mathclap{u \mid v \leadsto u}} d_{vu},$$ and the total distance to the inbound reachability set of $v$ as $$\overleftarrow{S}[v] = \sum_{\mathclap{u \mid u \leadsto v}} d_{uv}.$$ The outbound and inbound centralities are accordingly defined as the (inverse of the) ratios$\overrightarrow{S}[v]/\overrightarrow{R}[v]$ and $\overleftarrow{S}[v]/\overleftarrow{R}[v]$.

Unfortunately, the hybrid estimator, and even the special case of the pivoting estimator, do not work well with direction. This is because directed distances are not a metric (they are not symmetric). Intuitively, distances from the pivot (closest sampled node) can be much larger than distances from the node for which we estimate centrality.

Sampling can be used with direction, but, when naively applied, will not provide relative error guarantees even when the distance distribution is not skewed. The reason is that it is not enough to use all distances from a small sample of nodes. For sampling to work, we need to obtain a sample of a certain size from the reachability set of each node. Some nodes, however, may reach few or no nodes from this sample. Therefore the sample provides very little information (or none at all) for estimating the centrality of these nodes.

\begin{algorithm}[t]\caption{Estimate for all $v\in V$ average distance to reachable nodes $\hat{B}$ and cardinality $\hat{R}$: directed graphs \label{directed:alg}} \begin{algorithmic} \State $t\gets 0$, \For {$v\in V$} $\text{\sc mark}[v]\gets \textbf{False}$; $\text{{\sc count}}[v]\gets 0$; $\text{\sc T[v]}\gets 0$; $\text{\sc distsum}[v]\gets 0$ \EndFor \For {nodes $u\in V$ in random order} \State $t\gets t+1$; $\text{\sc mark}[u]\gets \textbf{True}$ \State Perform pruned Dijkstra from $u$ on $G^T$ \For {each scanned node $v$ of distance $d_{vu}$} \If {$\text{\sc count}[v]=k$} Prune Dijkstra at $v$ \Else \If {$u\not= v$ } \State $\text{\sc distsum}[v] \overset{+}{\gets} d_{vu}$ \State $\text{\sc count}[v] \overset{+}{\gets} 1$ \If {$\text{\sc count}[v]=k$} \State $\text{\sc T}[v]\gets t$ \If {$\text{\sc mark}[v]$} $\text{\sc T}[v]\gets t-1$ \EndIf \EndIf \EndIf \EndIf \EndFor \EndFor \For {$v\in V$} \If {$\text{\sc count}[v]=0$} $\hat{B}[v] \gets 0$ \Else $\, \hat{B}[v] \gets \text{\sc distsum}[v]/\text{\sc count}[v]$ \EndIf \If {$\text{\sc count}[v]<k$} $\hat{R}[v] \gets \text{\sc count}[v]$ \Else $\, \hat{R}[v] \gets 1+\frac{(k-1)(n-2)}{T[v]-1}$ \EndIf \EndFor \end{algorithmic} \end{algorithm}

We extend the basic sampling approach to directed graphs using an algorithm of Cohen~\cite{ECohen6f} that efficiently computes for each node a uniform sample of size $k$ from its reachability set~(for outbound centrality) or from nodes that can reach it (for inbound centrality). We modify the algorithm so that respective distances are computed as well. (We apply Dijkstra's algorithm instead of generic graph searches.) This algorithm also computes $nk$ distinct distances, but does so adaptively, so that they are not all from the same set of sources.

The same algorithm also provides approximate cardinalities of these sets~\cite{ECohen6f}. This means that, when the distance distribution is not too skewed, we can obtain good estimates of the average distance to reachable nodes (or from nodes our node is reachable from).

 Algorithm~\ref{directed:alg} contains pseudocode for estimating outbound average distance ($\overrightarrow{B} = \overrightarrow{S}/\overrightarrow{R}$) and reachability ($\overrightarrow{R}$) for all nodes. By applying the same algorithm on $G$ instead of the reverse graph~$G^T$, we can obtain estimates for the inbound quantities.

The algorithm computes for each node a uniform random sample of size $k$ from its reachability set. It does so by running Dijkstra's algorithm from each node $u$ in random order, adding~$u$ to the sample of all nodes it reaches. Since these searches are pruned at nodes whose samples already have $k$ nodes, no node is scanned more than $k$ times during the entire computation. The total cost is thus comparable to $k$ full (unpruned) Dijkstra computations. This algorithm does not offer worst-case guarantees. However, on realistic instances, where centrality is in the order of the median distance, it performs well.

\begin{algorithm}[t]\caption{Estimate for all $v\in V$ weighted sum of distances to reachable nodes $\hat{S}$ and weighted sum of reachable nodes $\hat{R}$: directed graphs \label{wdirected:alg}} \begin{algorithmic} \For {$v\in V$} $\text{{\sc count}}[v]\gets 0$; $\text{{\sc Bcount}}[v]\gets 0$; $\text{\sc distsum}[v]\gets 0$ \EndFor \State $V_+ \gets \{v\in V \mid \beta[v]>0 \}$ \For {$u\in V_+$} $r[v] \gets \text{\sc rand()}/\beta[v]$ \Comment{{\sc rand}$() \sim U[0,1]$ is uniform at random from $[0,1]$} \EndFor \For {$u\in V_+$ in increasing $r$ order} \State Perform pruned Dijkstra from $u$ on $G^T$ \For {each scanned node $v$ of distance $d_{vu}$} \If {$\text{\sc count}[v]=k$} Prune Dijkstra at $v$ \Else \If {$u\not= v$ } \State $\text{\sc count}[v] \overset{+}{\gets} 1$ \If {$\text{\sc count}[v] < k$} \State $\text{\sc distsum}[v] \overset{+}{\gets} \beta[u] d_{vu}$ \State $\text{\sc Bcount}[v] \overset{+}{\gets}\beta[u]$ \EndIf \If {$\text{\sc count}[v]=k$} \State $\text{\sc T}[v]= r[u]$ \EndIf \EndIf \EndIf \EndFor \EndFor \For {$v\in V$} \If {$\text{\sc count}[v]=0$} $\hat{R}[v] \gets 0$; $\hat{S}[v] \gets 0$ \ElsIf {$\text{\sc count}[v]< k$} $\hat{R}[v] \gets \text{\sc Bcount}[v]$; $\hat{S}[v] \gets \text{\sc distsum}[v]$ \Else $\, \hat{S}[v] \gets \frac{\text{\sc distsum}}{T[v]}$; $\hat{R}[v] \gets \frac{k-1}{T[v]}$ \EndIf \EndFor \end{algorithmic} \end{algorithm}

The algorithm applies a bottom-$k$ variant~\cite{bottomk07:ds} of the reachability estimation algorithm of Cohen~\cite{ECohen6f} and also computes distances. The cardinality estimator is unbiased with coefficient of variation~(CV) at most $1/\sqrt{k-2}$~\cite{ECohen6f}. The quality of the average distance estimates depends on the distribution of distances and we evaluate it experimentally.

\begin{table*} \centering \caption{Evaluating algorithms on \emph{undirected} instances. For each instance, we report its number of nodes and edges, and for several algorithms the running time and average relative error.} \label{tab:main} \begin{tabular}{@{}llrrrrrrrrrrr@{}} \toprule & & & & \textbf{Exact} & \multicolumn{2}{c}{\textbf{Sampling}} & \multicolumn{2}{c}{\textbf{Pivoting}} & \multicolumn{2}{c}{\textbf{Hyb.-0.1}} & \multicolumn{2}{c}{\textbf{Hyb.-ad}} \\ \cmidrule(lr){5-5}\cmidrule(lr){6-7}\cmidrule(lr){8-9}\cmidrule(lr){10-11}\cmidrule(l){12-13} & & $|V|$ & $|E|$ & time & err. & time & err. & time & err. & time & err. & time\\ type & instance & [$\cdot 10^3$] & [$\cdot 10^3$] & $\approx$\,[h:m] & [\%] & [sec] & [\%] & [sec] & [\%] & [sec] & [\%] & [sec] \\\midrule \instance{road} & \instance{fla-t} & 1\,070 & 1\,344 & 59:30 & 5.4 & 24.4 & 3.2 & 21.6 & 2.5 & 28.3 & 2.8 & 73.2\\ & \instance{usa-t} & 23\,947 & 28\,854 & 44\,222:06 & 2.9 & 849.4 & 3.7 & 736.4 & 2.0 & 2\,344.3 & 2.6 & 9\,937.9\\ \instance{grid} & \instance{grid20} & 1\,049 & 2\,095 & 70:34 & 4.3 & 26.5 & 3.5 & 26.8 & 2.9 & 29.2 & 3.3 & 69.7\\ \instance{triang} & \instance{buddha} & 544 & 1\,631 & 19:07 & 3.6 & 14.5 & 3.3 & 13.6 & 2.4 & 15.9 & 3.2 & 30.7\\ & \instance{buddha-w} & 544 & 1\,631 & 21:25 & 3.5 & 16.4 & 2.6 & 15.5 & 2.2 & 18.5 & 2.9 & 38.1\\ & \instance{del20-w} & 1\,049 & 3\,146 & 72:06 & 2.7 & 27.4 & 3.6 & 26.7 & 2.6 & 32.6 & 2.7 & 71.0\\ & \instance{del20} & 1\,049 & 3\,146 & 67:54 & 4.1 & 25.6 & 5.3 & 25.2 & 3.7 & 27.0 & 3.6 & 54.7\\ \instance{game} & \instance{FrozenSea} & 753 & 2\,882 & 38:25 & 3.0 & 22.1 & 4.1 & 20.2 & 2.1 & 24.0 & 3.4 & 49.3\\ \instance{sensor} & \instance{rgg20} & 1\,049 & 6\,894 & 137:36 & 1.6 & 54.2 & 3.8 & 49.3 & 2.1 & 63.7 & 2.2 & 123.3\\ & \instance{rgg20-w} & 1\,049 & 6\,894 & 160:29 & 1.6 & 61.2 & 3.8 & 57.1 & 2.1 & 73.3 & 2.3 & 142.3\\ \instance{comp} & \instance{Skitter} & 1\,695 & 11\,094 & 248:27 & 0.7 & 59.7 & 14.3 & 55.2 & 0.7 & 61.6 & 3.6 & 109.5\\ & \instance{MetroSec} & 2\,250 & 21\,643 & 269:51 & 0.6 & 52.1 & 2.3 & 47.5 & 0.6 & 53.2 & 0.3 & 93.2\\ \instance{social} & \instance{rws20} & 1\,049 & 3\,146 & 113:40 & 0.9 & 45.6 & 3.0 & 41.3 & 0.9 & 49.4 & 0.9 & 98.6\\ & \instance{rba20} & 1\,049 & 6\,291 & 132:35 & 0.8 & 56.8 & 9.7 & 48.4 & 0.8 & 60.2 & 1.0 & 117.4\\ & \instance{Hollywood} & 1\,069 & 56\,307 & 226:42 & 1.0 & 86.5 & 14.6 & 81.8 & 1.0 & 85.7 & 1.9 & 117.6\\ & \instance{Orkut} & 3\,072 & 117\,185 & 2\,973:09 & 1.7 & 377.4 & 7.2 & 367.6 & 1.7 & 376.4 & 2.1 & 553.0\\ \bottomrule \end{tabular} \end{table*}

We also consider non-uniform node weights and the respective weighted definitions, $\overrightarrow{S}_\beta[v] = \sum_{u | v \leadsto u} \beta(u) d_{vu}$ and $\overrightarrow{R}_\beta[v] = \sum_{u \mid v \leadsto u} \beta(u)$. A pseudocode for a weighted version is provided as Algorithm~\ref{wdirected:alg}. The algorithm assigns nodes with ranks that depend on their weight, effectively having each node count for a bottom-$k$ sample of its reachability set, as proposed by Cohen and Kaplan~\cite{ECohen6f,bottomk07:ds}. The pseudocode uses priority sampling \cite{Ohlsson_SPS:1998,DLT:jacm07}. The algorithm then processes nodes according to increasing rank order. The weighted reachability estimate is applied to the rank of the $k$th sample (this is a bottom-$k$ estimator).

\section{Related Work} \label{sec:related} Closeness centrality is only one of several common definitions of importance rankings. These include degree centrality, intended to capture activity level, betweenness centrality, which captures power, and eigenvalue centralities, which capture reputation \cite{Freeman:sn1979,wassermansocialnetworksbook}. We only consider the classic definition of closeness centrality. A well-studied alternative is {\em distance-decay} closeness centrality, where the contribution of each node to the centrality of another is discounted (is non-increasing) with distance \cite{Dangalchev:2006,CoKa:jcss07,Opsahl:2010,BoldiVigna:IM2014,BoldiVignaHyperball:arxiv2014,ECohenADS:PODS2014}. The subtle difference between distance-decay and classic closeness centrality is that the latter emphasizes the penalties for far nodes, whereas the distance-decay measures instead emphasize the reward from closer nodes. Distance-decay centrality is well defined on disconnected or directed graphs. In terms of scalable computation, efficient algorithms with a small relative error guarantee were known for two decades and engineered to handle graphs with billions of edges \cite{ECohen6f,CoKa:jcss07, PGF_ANF:KDD2002,CGLM:ICSR2011,hyperANF:www2011,BoldiRV11:socinfo2012,BackstromBRUV12:websci2012,ECohenADS:PODS2014}. These algorithms, however, provide no guarantees for estimating classic closeness centrality. The intuitive reason is that they are based on sampling that is biased towards closer nodes, whereas correctly estimating classic closeness centrality requires accounting for distant nodes, which can be missed by such a sample.

\section{Experiments} \label{experiments:sec}

We implemented our algorithms in C++ using Visual Studio~2013 with full optimization. We conducted all tests on a machine with two Intel Xeon E5-2690 CPUs and 384\,GiB of DDR3-1066 RAM, running Windows 2008R2 Server. Each CPU has 8~cores~(2.90\,GHz, 8\,$\times$\,64\,kiB L1, 8~$\times$\,256\,kiB, and 20\,MiB L3 cache), but all runs are sequential. We use 32-bit integers to represent arc lengths.

We test a variety of instances, including \emph{social networks} (\instance{Epinions}~\cite{rad-tmsw-03}, \instance{WikiTalk}~\cite{lhk-snsm-10,lhk-ppnlo-10}, \instance{Flickr}~\cite{mmgdb-maosn-07}, \instance{Hollywood} \cite{brsv-llpmc-11,bv-twfct-04}, \instance{Twitter}~\cite{TwitterData}, \instance{LiveJournal}~\cite{lldm-cslnn-09}, and \instance{Orkut}~\cite{OrkutData}), \emph{computer networks} (\instance{Gnutella}~\cite{rfi-mgnpl-02}, \instance{Skitter}~\cite{lkf-gtdls-05}, \instance{Slashdot} \cite{lldm-cslnn-09}, \instance{MetroSec}~\cite{mlh-fcetb-09}), and \emph{web graphs} (\instance{NotreDame}~\cite{ajb-dwww-99}, \instance{In\-do} \cite{brsv-llpmc-11,bv-twfct-04}, \instance{Indochina}~\cite{brsv-llpmc-11,bv-twfct-04}). All these instances are unweighted, and some are directed. We consider two additional synthetic instances: \instance{rws20} is generated according to a preferential attachment model~\cite{watts1998collective} and \instance{rba20} is a small-world graph~\cite{barabasi1999emergence}.

We also test \emph{road networks}~\cite{dgj-spndi-09}. Instances \instance{fla-t} (Florida) and \instance{usa-t} (USA) are undirected and use TIGER data; \instance{eur-t} and \instance{eur-d} are directed and represent Western Europe. For these instances, the suffix indicates whether edge costs represent travel times (\instance{-t}) or distances (\instance{-d}). Instance \instance{grid20} is a $1024 \times 1024$ unweighted grid.

\begin{figure*}\foreach \x in {3,5,7,8,10,11}{\includegraphics[width=0.33\textwidth,page=\x]{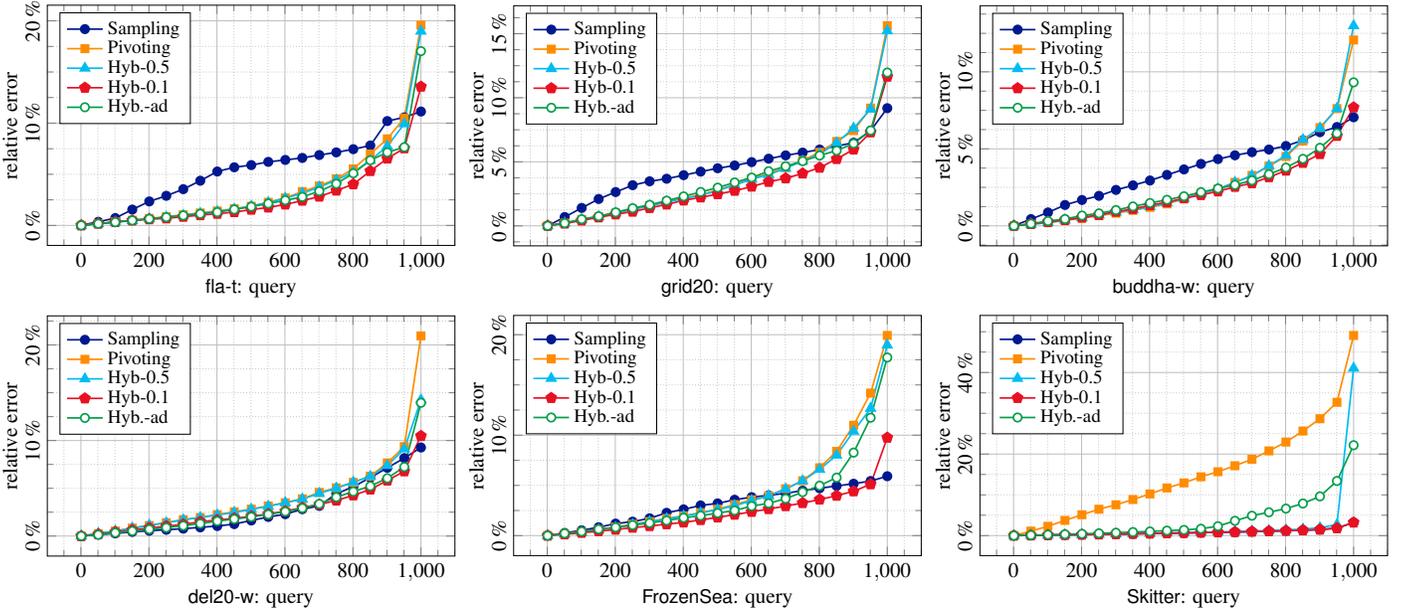} } \caption{Cumulative quality distribution (over 1000 queries) for varying $\epsilon$.}\label{fig:epsilon}\end{figure*}

The \instance{buddha} instance is a computer graphics mesh representing a three-dimensional object~\cite{SNCH-08}. Instance~\instance{del20} is a Delaunay triangulation of $2^{20}$ random points on the unit square~\cite{hss-eshqg-10}. Nodes also represent random points in the unit square for \instance{rgg20}, but now two nodes are connected by an edge if the corresponding Euclidean distance is below a given threshold~(chosen to ensure the graphs are almost connected \cite{hss-eshqg-10}). Such \emph{random geometric graphs} often model sensor networks. These three instances are unweighted; their counterparts with a \instance{-w} suffix have edge lengths corresponding to Euclidean distances. Instance~\instance{FrozenSea} is a grid with obstacles from Starcraft (a computer game) available from \url{movingai.com}~\cite{s-bgbp-12}. Edge lengths are set to~$408$ for axis-aligned moves and~$577$ for diagonal moves ($577/408 \approx \sqrt{2}$).

\subsection{Undirected Closeness Centrality}

Table~\ref{tab:main} summarizes the main results for undirected instances. We set $k = 100$ for this experiment. We evaluate sampling, pivoting, and our novel hybrid algorithm with respect to running time and solution quality. We consider two versions of our algorithm, both based on Algorithm~\ref{bavelasu:alg}: the first uses $\epsilon = \sqrt{1/k} = 0.1$; the \emph{adaptive} version picks, for each node, the $\epsilon$ value from \{0.001, 0.025, 0.05, 0.1, 0.2, 0.5, 0.99\} that minimizes the estimated error.

For each instance, Table~\ref{tab:main} shows the number of nodes and edges it contains~(in thousands), followed by the estimated time needed to compute exact centralities for all nodes. Then, for each approximate algorithm, we show its average relative error~(over 1000 random nodes queried) and the total time for computing centrality estimates for all nodes~(including preprocessing).

We observe that the exact algorithm is prohibitively time-consuming for large graphs, justifying our settling for approximations. Among those, all methods do reasonably well, with average relative error always below 15\%. The sampling algorithm is in general more robust than pivoting, with average relative error below 6\%. For some high-diameter graphs (such as road networks and meshes), however, pivoting finds better results. Our hybrid algorithm successfully achieves a good tradeoff between these two approaches. Its quality usually matches the best among pivoting and sampling, and often outperforms them.

The adaptive version of our algorithm goes one step further and actually uses different values of $\epsilon$ to obtain even finer tradeoffs. This can occasionally be helpful (as in \instance{MetroSec}), but in general using fixed $\epsilon$ is better in terms of running time and quality. Although Algorithm~\ref{bavelaserr:alg} uses additional space to make even finer choices, it leads to very similar results (not shown in the table). We conclude that fixing $\epsilon = \sqrt{1/k}$ is a good strategy: It is more robust than either sampling or pivoting, with very little overhead. On the biggest graph we tested (\instance{Orkut}), with 117 million edges, we obtained centrality estimates with approximation guarantees for all nodes in about six minutes.

Figure~\ref{fig:epsilon} examines the quality of the algorithms in Table~\ref{tab:main} in more detail. For comparison, we also show results for the hybrid algorithm with $\epsilon = 0.5$. Once again, we compute the relative error for 1000 queries, plotted in order of increasing error. In other words, for each value $1 \leq i \leq 1000$, we report the $i$-th smallest relative error observed for each algorithm. We consider six representative instances. For \instance{fla-t}, \instance{grid20}, and \instance{buddha-w}, sampling yields better results than pivoting; for \instance{del20-w}, \instance{FrozenSea}, and \instance{Skitter}, sampling behaves better. On all cases, our default hybrid algorithm (with $\epsilon = 0.1$) is generally better than either method. We note that, unsurprisingly, pivoting tends to have more outliers than pure sampling (i.e., the worst queries for pivoting are worse than the worst for sampling). Although some of this effect is transferred to the hybrid algorithm, it is much less pronounced. This is not true with higher $\epsilon$, which causes the hybrid algorithm to rely more heavily on pivoting.

\begin{table} \setlength{\tabcolsep}{1.25ex} \centering \caption{Evaluating algorithms on \emph{directed} instances. As in Table~\ref{tab:main}, we report the number of nodes and directed edges and for several algorithms the running time and average relative error.} \label{tab:directed} \begin{tabular}{@{}llrrrrr@{}} \toprule & & & & \textbf{Exact} & \multicolumn{2}{c}{\textbf{Sampling}}\\ \cmidrule(lr){5-5}\cmidrule(lr){6-7} & & $|V|$ & $|E|$ & time & err. & time\\ type & instance & [$\cdot 10^3$] & [$\cdot 10^3$] & $\approx$\,[h:m] & [\%] & [sec] \\\midrule \instance{road} & \instance{eur-t} & 18\,010 & 42\,189 & 28\,399:47 & 3.2 & 655.9\\ & \instance{eur-d} & 18\,010 & 42\,189 & 22\,306:20 & 3.2 & 517.0\\ \instance{web} & \instance{NotreDame} & 326 & 1\,470 & 0:54 & 2.4 & 1.5\\ & \instance{Indo} & 1\,383 & 16\,540 & 58:46 & 4.1 & 21.1\\ & \instance{Indochina} & 7\,415 & 191\,607 & 2\,884:19 & 4.7 & 174.7\\ \instance{comp} & \instance{Gnutella} & 63 & 148 & 0:02 & 2.8 & 0.6\\ \instance{social} & \instance{Epinions} & 76 & 509 & 0:07 & 5.4 & 1.1\\ & \instance{Slashdot} & 82 & 870 & 0:18 & 2.2 & 2.2\\ & \instance{Flickr} & 1\,861 & 22\,614 & 227:01 & 4.3 & 65.1\\ & \instance{WikiTalk} & 2\,394 & 5\,021 & 22:01 & 0.5 & 5.4\\ & \instance{Twitter} & 457 & 14\,856 & 28:16 & 1.2 & 26.1\\ & \instance{LiveJournal} & 4\,848 & 68\,475 & 2\,757:01 & 1.9 & 276.8\\ \bottomrule \end{tabular} \end{table}

\subsection{Directed Centrality}

We now consider centrality on arbitrary directed graphs. Table~\ref{tab:directed} gives the results obtained by Algorithm~\ref{directed:alg}. Once again, we use~$k= 100$ and evaluate the algorithm with 1000 random queries. The ``Exact'' column shows the estimated time for computing all $n$ outbound centralities using Dijkstra computations. We then show the average relative error (over the 1000 random queries) and the total running time to compute all $n$ centralities using Algorithm~\ref{directed:alg}. Although this algorithm has no theoretical guarantees, its average relative error is consistently below 6\% in practice. Moreover, it is quite practical, taking less than three minutes even on a graph with almost 200 million edges.

\section{Conclusion}

We presented a comprehensive solution to the problem of approximating, within a small relative error, the classic closeness centrality of all nodes in a network. We proposed the first near-linear-time algorithm with theoretical guarantees and provide a scalable implementation. Our experimental analysis demonstrates the effectiveness of our solution.

Our basic design and analysis apply in any metric space: Given the set of distances from a small random sample of the nodes to all other nodes, we can estimate, for each node, its average distance to all other nodes, with a small relative error. We therefore expect our estimators to have further applications.


\begin{thebibliography}{10}

\bibitem{ajb-dwww-99} R.~Albert, H.~Jeong, and A.-L. Barab{\'a}si. \newblock Internet: Diameter of the {W}orld-{W}ide {W}eb. \newblock {\em Nature}, 401:130--131, September 1999.

\bibitem{BackstromBRUV12:websci2012} L.~Backstrom, P.~Boldi, M.~Rosa, J.~Ugander, and S.~Vigna. \newblock Four degrees of separation. \newblock In {\em WebSci}, pp. 33--42, 2012.

\bibitem{barabasi1999emergence} A.-L. Barab{\'a}si and R.~Albert. \newblock Emergence of scaling in random networks. \newblock {\em Science}, 286(5439):509--512, 1999.

\bibitem{Bavelas:HumanOrg1948} A.~Bavelas. \newblock A mathematical model for small group structures. \newblock {\em Human Organization}, 7:16--30, 1948.

\bibitem{Bavelas:Acous1950} A.~Bavelas. \newblock Communication patterns in task oriented groups. \newblock {\em Journal of the Acoustical Society of America}, 22:271--282, 1950.

\bibitem{Beauchamp:BS1965} M.~A. Beauchamp. \newblock An improved index of centrality. \newblock {\em Behavioral Science}, 10:161--163, 1965.

\bibitem{brsv-llpmc-11} P.~Boldi, M.~Rosa, M.~Santini, and S.~Vigna. \newblock Layered label propagation: A multiresolution coordinate-free ordering for compressing social networks. \newblock In {\em Proceedings of the 20th international conference on World Wide Web}, pp. 587--596. 2011.

\bibitem{hyperANF:www2011} P.~Boldi, M.~Rosa, and S.~Vigna. \newblock {HyperANF}: Approximating the neighbourhood function of very large graphs on a budget. \newblock In {\em WWW}, 2011.

\bibitem{BoldiRV11:socinfo2012} P.~Boldi, M.~Rosa, and S.~Vigna. \newblock Robustness of social networks: Comparative results based on distance distributions. \newblock In {\em SocInfo}, pp. 8--21, 2011.

\bibitem{bv-twfct-04} P.~Boldi and S.~Vigna. \newblock The {WebGraph} framework {I}: {C}ompression techniques. \newblock In {\em Proceedings of the 13th international conference on World Wide Web}, pp. 595--602. 2004.

\bibitem{BoldiVignaHyperball:arxiv2014} P.~Boldi and S.~Vigna. \newblock In-core computation of geometric centralities with hyperball: A hundred billion nodes and beyond. \newblock In {\em ICDM workshops}, 2013. \newblock \url{http://arxiv.org/abs/1308.2144}.

\bibitem{BoldiVigna:IM2014} P.~Boldi and S.~Vigna. \newblock Axioms for centrality. \newblock {\em Internet Mathematics}, 2014.

\bibitem{Cha82} M.~T. Chao. \newblock A general purpose unequal probability sampling plan. \newblock {\em Biometrika}, 69(3):653--656, 1982.

\bibitem{ECohen6f} E.~Cohen. \newblock Size-estimation framework with applications to transitive closure and reachability. \newblock {\em J. Comput. System Sci.}, 55:441--453, 1997.

\bibitem{Ecohen5f} E.~Cohen. \newblock Undirected shortest-paths in polylog time and near-linear work. \newblock {\em J. Assoc. Comput. Mach.}, 47:132--166, 2000. \newblock Extended version of a STOC 1994 paper.

\bibitem{ECohenADS:PODS2014} E.~Cohen. \newblock All-distances sketches, revisited: {HIP} estimators for massive graphs analysis. \newblock In {\em PODS}. ACM, 2014.

\bibitem{varopt_full:CDKLT10} E.~Cohen, N.~Duffield, C.~Lund, M.~Thorup, and H.~Kaplan. \newblock Efficient stream sampling for variance-optimal estimation of subset sums. \newblock {\em SIAM J. Comput.}, 40(5), 2011.

\bibitem{CoKa:jcss07} E.~Cohen and H.~Kaplan. \newblock Spatially-decaying aggregation over a network: {M}odel and algorithms. \newblock {\em J. Comput. System Sci.}, 73:265--288, 2007. \newblock Full version of a SIGMOD 2004 paper.

\bibitem{bottomk07:ds} E.~Cohen and H.~Kaplan. \newblock Summarizing data using bottom-k sketches. \newblock In {\em ACM PODC}, 2007.

\bibitem{CGLM:ICSR2011} P.~Crescenzi, R.~Grossi, L.~Lanzi, and A.~Marino. \newblock A comparison of three algorithms for approximating the distance distribution in real-world graphs. \newblock In {\em TAPAS}, 2011.

\bibitem{Dangalchev:2006} C.~Dangalchev. \newblock Residual closeness in networks. \newblock {\em Phisica A}, 365, 2006.

\bibitem{TwitterData} M.~De~Domenico, A.~Lima, P.~Mougel, and M.~Musolesi. \newblock The anatomy of a scientific rumor. \newblock {\em Scientific Reports}, 3:2980, 2013.

\bibitem{dgj-spndi-09} C.~Demetrescu, A.~V. Goldberg, and D.~S. Johnson, editors. \newblock {\em The Shortest Path Problem: Ninth {DIMACS} Implementation Challenge}, DIMACS Book~74. \newblock American Mathematical Society, 2009.

\bibitem{DLT:jacm07} N.~Duffield, M.~Thorup, and C.~Lund. \newblock Priority sampling for estimating arbitrary subset sums. \newblock {\em J. Assoc. Comput. Mach.}, 54(6), 2007.

\bibitem{EW_centrality:SODA2001} D.~Eppstein and J.~Wang. \newblock Fast approximation of centrality. \newblock In {\em SODA}, pp. 228--229, 2001.

\bibitem{Freeman:sociometry1977} L.~C. Freeman. \newblock A set of measures of centrality based on betweeness. \newblock {\em Sociometry}, 40:35--41, 1977.

\bibitem{Freeman:sn1979} L.~C. Freeman. \newblock Centrality in social networks: Conceptual clarification. \newblock {\em Social Networks}, 1, 1979.

\bibitem{hss-eshqg-10} M.~Holtgrewe, P.~Sanders, and C.~Schulz. \newblock Engineering a scalable high quality graph partitioner. \newblock In {\em 24th International Parallel and Distributed Processing Symposium (IPDPS'10)}, pp. 1--12. IEEE Computer Society, 2010.

\bibitem{Indyk:stoc1999} P.~Indyk. \newblock Sublinear time algorithms for metric space problems. \newblock In {\em STOC}. ACM, 1999.

\bibitem{Knuth2f} D.~E. Knuth. \newblock {\em The Art of Computer Programming, Vol 2, Seminumerical Algorithms}. \newblock Addison-Wesley, 1st edition, 1968.

\bibitem{lhk-ppnlo-10} J.~Leskovec, D.~Huttenlocher, and J.~Kleinberg. \newblock Predicting positive and negative links in online social networks. \newblock In {\em Proceedings of the 19th international conference on World wide web}, pp. 641--650. ACM, 2010.

\bibitem{lhk-snsm-10} J.~Leskovec, D.~Huttenlocher, and J.~Kleinberg. \newblock Signed networks in social media. \newblock In {\em Proceedings of the SIGCHI Conference on Human Factors in Computing Systems}, pp. 1361--1370. ACM, 2010.

\bibitem{lkf-gtdls-05} J.~Leskovec, J.~Kleinberg, and C.~Faloutsos. \newblock Graphs over time: {D}ensification laws, shrinking diameters and possible explanations. \newblock In {\em Proceedings of the eleventh ACM SIGKDD international conference on Knowledge discovery in data mining}, pp. 177--187. ACM, 2005.

\bibitem{lldm-cslnn-09} J.~Leskovec, K.~J. Lang, A.~Dasgupta, and M.~W. Mahoney. \newblock Community structure in large networks: Natural cluster sizes and the absence of large well-defined clusters. \newblock {\em Internet Mathematics}, 6(1):29--123, 2009.

\bibitem{Linsocial:book} N.~Lin. \newblock {\em Foundations of Social Search}. \newblock McGraw-Hill Book Co., New York, 1976.

\bibitem{mlh-fcetb-09} C.~Magnien, M.~Latapy, and M.~Habib. \newblock Fast computation of empirically tight bounds for the diameter of massive graphs. \newblock {\em Journal of Experimental Algorithmics (JEA)}, 13:10:1--10:9, 2009.

\bibitem{rfi-mgnpl-02} R.~Matei, A.~Iamnitchi, and I.~Foster. \newblock Mapping the {G}nutella network: Properties of large-scale peer-to-peer systems and implications for system design. \newblock {\em IEEE Internet Computing Journal}, 2002.

\bibitem{mmgdb-maosn-07} A.~Mislove, M.~Marcon, K.~P. Gummadi, P.~Druschel, and B.~Bhattacharjee. \newblock Measurement and analysis of online social networks. \newblock In {\em Proceedings of the 7th ACM SIGCOMM conference on Internet measurement}, pp. 29--42. 2007.

\bibitem{Ohlsson_SPS:1998} E.~Ohlsson. \newblock Sequential poisson sampling. \newblock {\em J. Official Statistics}, 14(2):149--162, 1998.

\vfill\eject

\bibitem{OkamotoCL:FAW2008} K.~Okamoto, W.~Chen, and X.~Li. \newblock Ranking of closeness centrality for large-scale social networks. \newblock In {\em Proc. 2nd Annual International Workshop on Frontiers in Algorithmics}, FAW. Springer-Verlag, 2008.

\bibitem{Opsahl:2010} T.~Opsahl, F.~Agneessens, and J.~Skvoretz. \newblock Node centrality in weighted networks: Generalizing degree and shortest paths. \newblock {\em Social Networks}, 32, 2010. \newblock \url{http://toreopsahl.com/2010/03/20/}.

\bibitem{PGF_ANF:KDD2002} C.~R. Palmer, P.~B. Gibbons, and C.~Faloutsos. \newblock {ANF:} {A} fast and scalable tool for data mining in massive graphs. \newblock In {\em KDD}, 2002.

\bibitem{rad-tmsw-03} M.~Richardson, R.~Agrawal, and P.~Domingos. \newblock Trust management for the semantic web. \newblock In {\em The Semantic Web -- ISWC 2003}, pp. 351--368. Springer, 2003.

\bibitem{Sabidussi:psychometrika1966} G.~Sabidussi. \newblock The centrality index of a graph. \newblock {\em Psychometrika}, 31(4):581--603, 1966.

\bibitem{SNCH-08} P.~V. Sander, D.~Nehab, E.~Chlamtac, and H.~Hoppe. \newblock Efficient traversal of mesh edges using adjacency primitives. \newblock {\em ACM Transactions on Graphics (TOG)}, 27(5):144, 2008.

\bibitem{s-bgbp-12} N.~R. Sturtevant. \newblock Benchmarks for grid-based pathfinding. \newblock {\em IEEE Transactions on Computational Intelligence and AI in Games}, 4(2):144--148, 2012.

\bibitem{Thorup:icalp2001} M.~Thorup. \newblock Quick $k$-median, $k$-center, and facility location for sparse graphs. \newblock In {\em ICALP}. Springer-Verlag, 2001.

\bibitem{UY} J.~D. Ullman and M.~Yannakakis. \newblock High-probability parallel transitive closure algorithms. \newblock {\em SIAM J. Comput.}, 20:100--125, 1991.

\bibitem{Vit85} J.~Vitter. \newblock Random sampling with a reservoir. \newblock {\em ACM Trans. Math. Softw.}, 11(1):37--57, 1985.

\bibitem{wassermansocialnetworksbook} S.~Wasserman and K.~Faust, editors. \newblock {\em Social Network Analysis: Methods and Applications}. \newblock Cambridge University Press, 1994.

\bibitem{watts1998collective} D.~J. Watts and S.~H. Strogatz. \newblock Collective dynamics of `small-world' networks. \newblock {\em Nature}, 393(6684):440--442, 1998.

\bibitem{OrkutData} J.~Yang and J.~Leskovec. \newblock Defining and evaluating network communities based on ground-truth. \newblock In {\em Proceedings of the ACM SIGKDD Workshop on Mining Data Semantics}, MDS '12, pp. 3:1--3:8, New York, NY, USA, 2012. ACM.

\end{thebibliography}
\end{document}